\documentclass[a4paper,10pt]{article} 
\usepackage{graphicx,amsfonts}

\begin{document}

\title{Nonstandard entropy production in the standard map}

%\author{
%   F. Baldovin\thanks{E-mail addresses: baldovin@cbpf.br, tsallis@cbpf.br},
%   C. Tsallis$^*$ and B. Schulze}
%\address{
%   Centro Brasileiro de Pesquisas F\'{\i}sicas\\
%   Rua Xavier Sigaud 150,22290-180 Rio de Janeiro -- RJ, Brazil}

%\author{A. Alpha\thanks{Corresponding author, alpha@omega.br}}
%\address{Alpha Institute}
%\author{B. Beta}
%\address{Beta University}

\author{F. Baldovin\thanks{E-mail address: baldovin@cbpf.br},
 C. Tsallis\thanks{E-mail address: tsallis@cbpf.br} and B. Schulze\\
\it{Centro Brasileiro de Pesquisas F\'{\i}sicas}\\
\it{Rua Xavier Sigaud 150,}\\
\it{Urca 22290-180 Rio De Janeiro -- RJ, Brazil } }
\date{March 23, 2002}

%\begin{document}

\maketitle

\begin{abstract}
We investigate the time evolution of the  entropy for a paradigmatic
conservative dynamical system, the standard map, for different values of its
controlling parameter $a$. When the phase space is sufficiently ``chaotic''
(i.e., for large $|a|$), we reproduce previous results. For
small values of $|a|$, when the phase space becomes an intricate structure with
the coexistence of chaotic and regular regions, an anomalous regime emerges.
We characterize this anomalous regime with  the generalized nonextensive entropy,
and we observe that for values of $a$ approaching zero, it lasts for
an increasingly large time.
This scenario displays a striking analogy with recent observations made in
isolated classical long-range $N$-body Hamiltonians, where, for a large class
of initial conditions, a metastable state (whose duration diverges with
$1/N\rightarrow 0$) is observed before it crosses over to the usual,
Boltzmann-Gibbs regime.
\end{abstract}

PACS numbers: 05.20.-y, 05.45.-a, 05.70.Ce

%\begin{cols}{2}
In his critical remarks about the domain of validity of Boltzmann principle,
Einstein stressed \cite{einstein} that the basis of statistical mechanics lies
on dynamics. During the past century an impressive amount of work has been
made to clarify this connection, but certainly this crucial point is
far from being well understood. In this paper, following recent 
observations on dissipative maps,
we analyze numerically the entropy production of a far-from-equilibrium
conservative dynamical system,
and we exhibit a new scenario where anomalous
effects arise, and that
may serve as a basis for extending the usual statistical-mechanical formalism.

In low-dimensional dynamical systems, both dissipative and conservative,
intensive effort has been done in analyzing the connection between
the property of mixing of the system and the evolution of the statistical entropy
(see, e.g., \cite{tsallis4,costa,lyra,tirnakli,moura,latora1,latora} and references
therein).
Particularly, an interesting observation was made in \cite{latora}, where
it was  found a simple connection between the Kolmogorov-Sinai entropy
rate (the one that stems from the properties of  mixing of the system),
and the statistical entropy (functional of a probability
distribution). It was shown in fact that,
partitioning the phase space
in $W$ cells
and starting from $M>>1$ points within one cell (details are given later),
the time dependence of the usual,
Boltzmann-Gibbs (BG), statistical entropy
\begin{equation}
 \label{SB}
S_1=-\sum_{i=1}^W p_{i}\ln{p_{i}},
\end{equation}
includes a {\em linear}
stage whose slope 
$K_1\equiv\lim_{t\to\infty}\lim_{W\to\infty}\lim_{M\to\infty}\frac{S_1(t)}{t}$
coincides with the Kolmogorov-Sinai entropy
rate, calculated for instance
via Pesin equality using the Lyapunov exponents.
Nevertheless, in the context of one-dimensional dissipative systems
(e.g., the logistic map), it was found in \cite{latora1} that
{\it at the edge of chaos} the usual
entropy (\ref{SB}) fails to exhibit with {\em nonvanishing} slope
such a behavior (which is of course expected since the Lyapunov exponent vanishes). 
Instead, the nonextensive entropy
\cite{tsallis} (for a recent review,  see
\cite{tsallis2})
\begin{equation}
\label{ST}
S_q=\frac{1-\sum_{i=1}^W p_{i}^{q} }{q-1}
\;\;\;(q\in\mathbb{R}),
\end{equation}
succeeds for
a specific value of $q< 1$ ($q^*\simeq 0.2445$ for the logistic map). We remind
that the entropic form (\ref{ST})
reduces to (\ref{SB}) in the limit $q\rightarrow 1$.

In Hamiltonian dynamics, one of the most studied
models is the {\em standard map} (see, for instance, \cite{tabor}), also
referred to as the {\em kicked-rotator} model:
\begin{eqnarray}
\label{std}
x_{t+1}&=&y_t+\frac{a}{2\pi}\sin(2\pi x_t)+x_t~~~\textup{(mod
1)},\nonumber\\ \\
y_{t+1}&=&y_t+\frac{a}{2\pi}\sin(2\pi x_t)~~~~~~~~~\textup{(mod 1)}\nonumber,
\end{eqnarray}
where $a\in\mathbb{R}$ ($t=0,1,2,...$). The system is integrable when $a=0$, while,
for large-enough values of $a$ (typically
$|a|\geq7$), it is strongly
chaotic\footnote {We are not interested here in the accelerator-mode islands
that appear at various critical values of $a$ (see, for instance, \cite{zaslavsky}
and references therein).}; in Fig. 1 we illustrate the
phase portrait of the map (\ref{std})
for different values of $a$.
It can be easily verified that, for all values of $a$, this two-dimensional
map is conservative; it has a pair of $(x,y)$-dependent Lyapunov exponents,
which only differ in sign (simplectic structure) and globally vanish
when $a$ vanishes.
Inspired by the results obtained in \cite{latora1,latora},
in this letter we study, for the standard map, the time
evolution of both the extensive ($q=1$) and the nonextensive
($q\neq 1$) statistical entropies, focusing our
attention on small values of the parameter $a$, i.e., on those situations where
the border between the chaotic and the regular regions has a relevant influence.
For smaller and smaller values of $|a|$ (say below $7$),
one encounters in fact an increasingly rich fractal-like structure,
characteristic of a large class of Hamiltonian systems. Typically, chains of
regular islands corresponding to elliptic points in resonance condition begin
to appear, and between each couple of elliptic points of a chain there is an
hyperbolic point, thus forming another chain of hyperbolic points responsible
for the chaotic areas. Around each
island there is another chain of islands of higher order and, once again,
another chain of hyperbolic points between the islands (see, for instance,
\cite{henon} for details). This structure is usually referred to as
``islands-around-islands''. If we think that in some sense the
islands-around-islands structure plays, for a Hamiltonian system, 
the role that edge-of-chaos plays for a dissipative system,
we verify that for the standard map its relative area
of influence increases when $a$
approaches $0$.

\begin{figure}
%\label{fig_1}
\begin{center}
\includegraphics[width=7cm,angle=0]{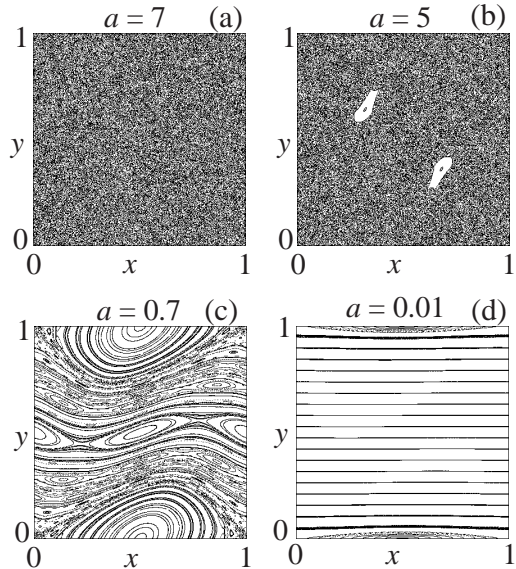}
\end{center}
\caption{\small Phase portrait of the standard map for
typical values of $a$. $M=20\times20$ orbits (black dots)
were started
with a uniform  distribution in the unit square 
and traced for $0\leq t\leq 200$.}
\end{figure}

To study numerically
the time evolution of the 
entropy, we introduce a coarse-graining
partition of the mapping phase space by dividing it in $W$ cells of equal
size, and we set many copies of the system ($M$ points) in a
far-from-equilibrium situation putting all the $M$ points randomly or
uniformly distributed inside a {\em single} cell.
The occupation number $M_i$ of
each cell $i$ ($\sum_{i=1}^W M_i=M$) provides a probability distribution
$p_i(t)\equiv M_i(t)/M$, hence an entropy value $S_q(t)\;(\forall q)$. Using then
the dynamic equations (\ref{std}), at each step the points spread in the
phase space causing the entropy value to change. In order to obtain a
consistent definition of the probabilities $\{p_i\}$ for all $t$ we are analyzing,
we impose the constraint $M>>W_{max}$, where $W_{max}\leq W$ is the maximum
number of cells occupied.
Finally, a {\em global} quantity over the whole $(x,y)$ phase
space is obtained averaging many histories. Each history
is characterized by the choice of the cell with which we start at $t=0$
(see \cite{baldovin_physica_A} for a preliminary exhibition of this phenomenon).
These cells are randomly chosen in the unit square.
We stress here that in this way
we are implementing a Gibbsian-like viewpoint in the sense that we
obtain an ensemble description of how the
system evolves towards equilibrium.
Given $a$, the result of this
analysis
is then a single curve of the entropy versus time, for each entropic
form $S_q$.
The initial part of the curve characterizes the entropy production
when the system is very far from equilibrium. Then comes an intermediate regime
whose duration increases with $W$. Finally, the entropy starts approaching its
equilibrium, where the entropy saturates at a value equal or below
$(W^{1-q}-1)/(1-q)$ (value of $S_q$ at equiprobability). 
With the aim of mathematically characterizing, for fixed $a$, the properties of the
entropy production in the limit $W\to\infty$, we 
perform the calculation for increasingly large
values of $W$. The curves asymptotically approach the $W\to\infty$ curve (see Fig. 2).
Numerically, the crucial point is to execute many times the calculation
with large $W$ satisfying the constraint $M>>W_{max}$. This exhausts
the memory capacity of a single, standard computer if we wish to attain $W$ larger than
say $10^6$; also, a lot of machine-time
is required. To overcome these problems,
we implemented a distributed computing technique \cite{baldovin}
which allows for the simultaneous utilization of many computers linked within a network.

\begin{figure}
%\label{fexp}
\begin{center}
\includegraphics[width=6cm,angle=0]{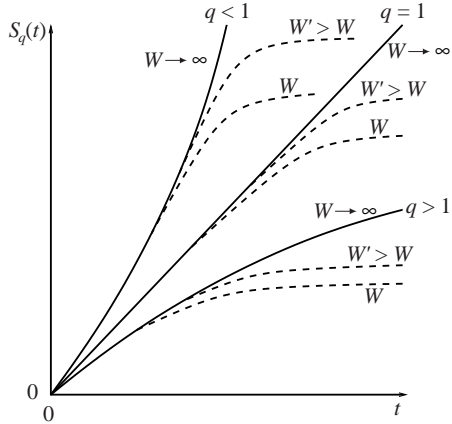}
\end{center}
\caption{\small Schematic representation of $S_q(t)$ for different
values of $q$, in the case of a strongly chaotic phase space.
Only for $q=1$ we have, in the limit $W\to\infty$,
a {\em finite} entropy production
$K_1$ ($K_q=0$ for $q>1$ and $K_q\to\infty$ for $q<1$).}
\end{figure}

Let us start with $a=7$, where the phase space is characterized
by a connected chaotic sea (Fig. 1(a)).
In this case there is just one value of the entropic index ($q=1$) that yields
a {\em linear} time evolution of the entropy $S_q(t)$,
after a possible transient that depends
on the details of the region where the initial data are set, and before the
effects of saturation. 
After the transient and before saturation, $S_q(t)$  is a convex function for $q<1$
and a concave function for $q>1$.
We schematize this analysis in Fig. 2, neglecting the initial transient.
Moreover, as said before,
during the linear stage, the slope of the BG entropy $S_1(t)$
coincides with the positive Lyapunov exponent.

%\newpage
\begin{figure}
%\label{figa5}
\begin{center}
\includegraphics[width=8cm,angle=0]{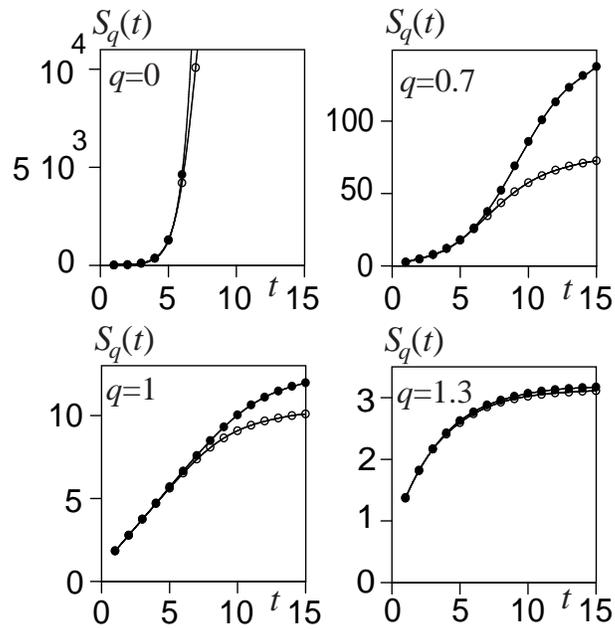}
\end{center}
\caption{\small $S_q(t)$ for $a=5$ and different values of $q$
($1000$ histories were averaged). Full (empty) circles correspond to $M/10=W=708\times 708$
($M/10=W=224\times 224$).
The lines are guides to the eye.
The slope of $S_1(t)$ between $t_1=2$ and $t_2=6$ is
$K_1(5)\simeq 0.97$.}
\end{figure}

If we now pass to $a=5$, the phase space presents
two large (and infinitely many small) islands and a connected chaotic sea (Fig. 1(b)).
Depending on where we set the initial data,
we may now face a strongly chaotic, or a regular, or even an
islands-around-islands region. Our averaging procedure does not
privilege any of these regions and allows the predominant ones to emerge. The
key point is to perform many
different histories so that the average stabilizes on a
definite curve. The resulting entropy curves reflect two
different phenomena: the sensitivity to the initial conditions of each area
and the relative extension of that area with respect to the whole phase space (unit square).
In spite of the presence of large islands, the results we obtain in this case 
qualitatively coincide with
Fig. 2; the only difference with the case $a=7$ is that the slope of the linear stage
of $S_1(t)$, $K_1(a)$, is smaller, consistently with the observation that the positive
Lyapunov exponent is approaching $0$.
In Fig. 3 we present the actual numerical results for $a=5$; our estimation of
 $K_1(5)\simeq 0.97$ is slightly smaller from
that found in \cite{latora} because here we average over the entire 
phase space, {\em including the islands}.

%\newpage
\begin{figure}
%\label{figa5}
\begin{center}
\includegraphics[width=8cm,angle=0]{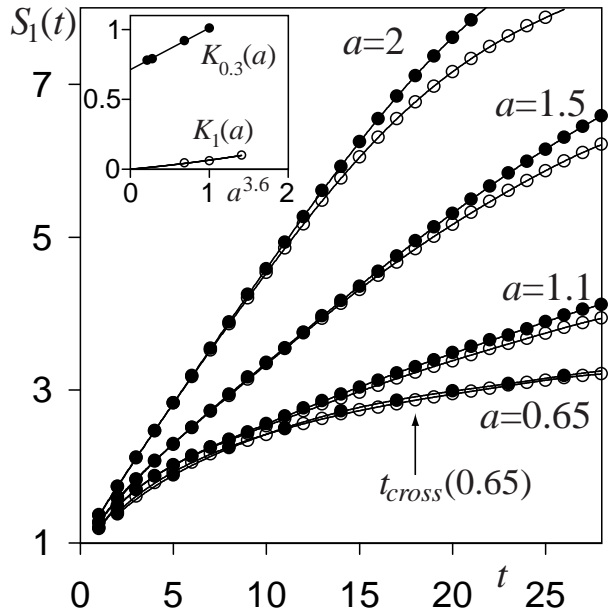}
\end{center}
\caption{\small $S_1(t)$ for different  values of $a\leq 2$
($5000$ to $7000$ histories were averaged). Full circles  correspond to $M=W=1000\times 1000$
for $a=2,1.5,1.1$
($M=W=5000\times 5000$ for $a=0.65$);
empty circles correspond to $M=W=448\times 448$ for $a=2,1.5,1.1$
($M=W=2236\times 2236$ for $a=0.65$).
Inset: Slopes of $S_1(t)$ ($K_1(a)$) and of $S_{0.3}(t)$  
($K_{0.3}\simeq 0.71+0.30 \;a^{3.6}$) 
in their linear regimes (see also Fig. 5).
The lines are guides to the eye.}
\end{figure}

What happens if we further reduce $a$, approaching situations similar to Fig. 1(c)?
Let us first concentrate on the BG entropy $S_1(t)$.
Fig. 4 displays that, for $a$ approaching zero,
the part which for large values of $a$ was linear starts bending downwards. 
Then, after a
certain characteristic time (noted $t_{cross}(a)$),
linearity is obtained once again.
If we now change $q$, we find that only for a special value $q^*\simeq 0.3$,
$S_{q^*}(t)$ grows linearly
for $t\ t_{cross}(a)$.
In Fig. 5 we plot $S_{0.3}(t)$ for $a$ approaching zero.
For small values of $a$ and for $t\leq t_{cross}$, 
the dynamical system appears to be trapped in islands-around-islands regions, 
so that the usual entropy, $S_1$, is forced to grow
slowly. For $t$ above $t_{cross}$, 
the normal dynamical behavior is restored;
in other words, there is a {\em crossover} from
an ``anomalous'' regime, where $S_{q^*}$ grows linearly, to the ``usual'' regime, where,
in accordance with the observation in \cite{latora}, it is
$S_1$ that grows linearly (before saturation).
It is relevant to notice that $t_{cross}$ diverges for $a\to 0$
(see inset of Fig. 5).
In order to clarify these complex phenomena, 
we have schematized them in Fig. 6.
In the ``usual'' regime a connection with an {\em exponential} sensitivity
to the initial condition is known; the connection of the ``anomalous''
regime to a {\em power-law} sensitivity has been preliminarily verified and is now under 
detailed investigation 
\cite{baldovin1} (see also \cite{tsallis4,costa,lyra,tirnakli}).

%\newpage
\begin{figure}
%\label{figa5}
\begin{center}
\includegraphics[width=7cm,angle=0]{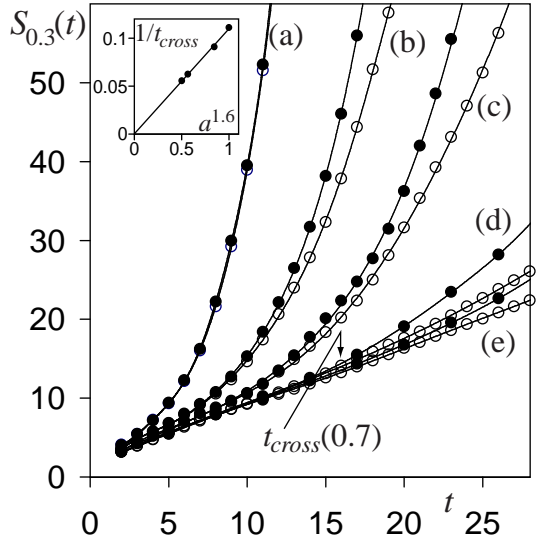}
\end{center}
\caption{\small $S_{0.3}(t)$ for: $a=1.5$ (a); $a=1.1$ (b);
$a=0.9$ (c); $a=0.7$ (d); $a=0.65$ (e)
($5000$ to $7000$ histories were averaged). Full circles  correspond to $M=W=1000\times 1000$
for (a), (b), (c),
$M=W=5000\times 5000$ for (d), (e);
empty circles correspond to $M=W=448\times 448$ for (a), (b), (c),
$M=W=2236\times 2236$ for (d), (e).
Inset: $t_{cross}(a)$ defined as the intersection of the linear part (before it starts bending)
and a standard extrapolation of the bended part of the curves $S_{0.3}(t)$. 
Notice that $t_{cross}(a)$
diverges for $a\to 0$.
The lines are guides to the eye.
Our results suggest 
$\lim_{t\to\infty}\lim_{a\to 0}\lim_{W\to\infty}\lim_{M\to\infty}\frac{S_{0.3}(t)}{t}\simeq 0.71$
for $q=q^*\simeq 0.3$, whereas this limit vanishes (diverges) for $q>q^*$ ($q<q^*$).}
\end{figure}

\begin{figure}
%\label{figpower}
\begin{center}
\includegraphics[width=6cm,angle=0]{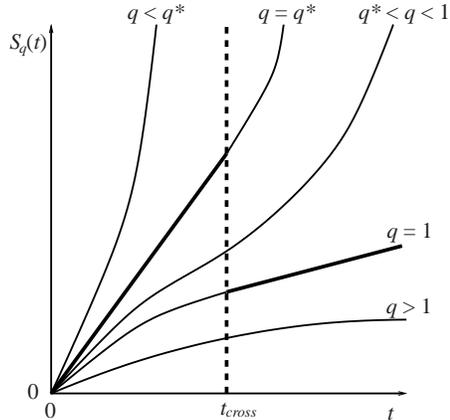}
\end{center}
\caption{\small Schematic representation of $S_q(t)$
for different values of $q$, in the limit $W\to\infty$,
when there is a crossover between two different regimes.}
\end{figure}

Summarizing, we have studied the production of entropy for a well known
low-dimensional conservative map controlled by the parameter $a$.
It appears that, for fixed $a\neq 0$ and in the limit $t \rightarrow \infty$
(to be in all cases taken {\it after} the $W \rightarrow \infty$ limit),
it is the standard, BG entropy $S_1$, which is
associated with a {\it finite} entropy production. However, during the
time interval preceding this extreme limit, an {\em increasingly large}
interval emerges for which it is $S_{q^*}$, with $q^* \simeq 0.3$, which
is associated with a {\it finite} entropy production. Consistently,
our results strongly suggest that,
in the $lim_{t \rightarrow \infty}\;lim_{a \rightarrow 0}$
ordering, the only {\it linear} regime in fact observed is that corresponding
to $q^* \simeq 0.3$, whereas in the $lim_{a \rightarrow 0}\;lim_{t \rightarrow \infty}$
ordering the {\it linear} regime observed corresponds to $q=1$. 
This fact provides a very appealing scenario
for (meta) equilibrium thermostatistics in long-range Hamiltonians
such as that described in \cite{latora2,campa}. For these systems, a
longstanding metastable state can exist (preceding the BG one)
whose duration diverges with $N \rightarrow \infty$, and whose distribution
of velocities is {\it not} Maxwellian, but rather a power-law. The role
played by $a$ in our present simple map
and that played by $1/N$ in such long-range-interacting many-body systems,
appear to be very similar, thus providing a
dynamical basis for nonextensive statistical mechanics
\cite{tsallis}.

\section{Acknowledgments}    
We acknowledge E.G.D. Cohen for stressing our attention on Einstein's 1910
paper. We have benefitted from partial support by PRONEX, CNPq, CAPES and
FAPERJ (Brazilian agencies).

%\end{cols}

\end{document}